\documentclass[12pt,preprint]{aastex}
\def\gs{\mathrel{\raise0.35ex\hbox{$\scriptstyle >$}\kern-0.6em
\lower0.40ex\hbox{{$\scriptstyle \sim$}}}}
\def\ls{\mathrel{\raise0.35ex\hbox{$\scriptstyle <$}\kern-0.6em
\lower0.40ex\hbox{{$\scriptstyle \sim$}}}}

\shorttitle{Faint CO J=6--5  in Arp\,220}
\shortauthors{Papadopoulos, Isaak, \& van der Werf}

\begin{document}

\title{CO J=6--5 in Arp\,220:  strong  effects of dust on high-J CO  lines}

\author{Padeli \ P.\ Papadopoulos}
\affil{Argelander-Institut f\"ur Astronomie,  Auf dem H\"ugel 71,  D-53121 Bonn, Germany}
\email{padeli@astro.uni-bonn.de}

\author{Kate Isaak}
\affil{School of Physics and Astronomy, University of Wales, Cardiff CF24 3YB, UK}
\email{Kate.Isaak@astro.cf.ac.uk}

\and

\author{Paul van der Werf}
\affil{Leiden Observatory, Leiden University, P.O. Box 9513, 2300 RA Leiden, The Netherlands}
\email{pvdwerf@strw.leidenuniv.nl}

\begin{abstract}

We  report  new  single  dish  CO J=6--5  line  observations  for  the
archetypal Ultra  Luminous Infrared  Galaxy (ULIRG) Arp\,220  with the
James Clerk  Maxwell Telescope atop  Mauna Kea in Hawaii.   The J=6--5
line  is  found  to  be  faint,  with  brightness  temperature  ratios
(6--5)/(1--0), (6--5)/(3--2)  of $\rm R_{65/10}$=0.080$\pm  $0.017 and
$\rm  R_{65/32}$=0.082$\pm  $0.019,  suggesting  very  low  excitation
conditions  that cannot  be reconciled  with the  warm and  very dense
molecular gas  present in  one of the  most extreme starbursts  in the
local Universe.  We find that  an optically thick dust continuum, with
$\rm \tau (\nu  \ga 350\,GHz)\ga 1$ for the bulk of  the warm dust and
gas in  Arp\,220, submerges this line  to an almost  black body curve,
reducing  its flux,  and {\it  affecting its  CO Spectral  Line Energy
Distribution  (SLED) at  high  frequencies}.  This  also resolves  the
C$^+$ line deficiency in this object, first observed by {\it ISO}: the
near absence of that line is  a dust optical depth effect, not a dense
Photodissociation Region (PDR) phenomenon.  Finally we briefly comment
on the possibility  of such extreme ISM states in  other ULIRGs in the
distant Universe,  and its consequences for the  diagnostic utility of
high frequency molecular and atomic ISM lines in such systems.  In the
case of Arp\,220 we anticipate  that the now spaceborne Herschel Space
Observatory will find  faint high-J CO lines at  $\nu $$\ga $ 690\,GHz
that would appear  as sub-thermally excited with respect  to the low-J
ones as a result of the effects of dust absorption.

\end{abstract}

\keywords{galaxies: individual (Arp\,220) --- galaxies: ISM --- 
galaxies: starburst --- ISM: molecules}

\section{Introduction}

Since   the   discovery   of   its  enormous   IR   luminosity   ($\rm
  L_{IR}$(8--1000\,$\mu      $m)$\sim     $$1.6\times     10^{12}$$\rm
  \,L_{\odot}$,  Soifer   et  al.    1984;  Emerson  et   al.   1984),
  representing $\sim 99\%$ of  its bolometric luminosity, Arp\,220 has
  been a  study of  extremes when  it comes to  the ISM  conditions in
  intense starbursts in  the local Universe.  A prominent  member in a
  prominent  class of  galaxies (e.g.   Sanders \&  Ishida  2004), the
  large  IR  luminosity  and  relative proximity  (at  $\rm  D_L$$\sim
  $77\,Mpc, it is the nearest ULIRG), made Arp\,220 an early target of
  numerous  molecular line  observations.   Its large  CO J=1--0  line
  luminosity  revealed  a  huge  molecular gas  reservoir  ($\rm  \sim
  10^{10}\,M_{\odot}$)  (Young  et al.   1984;  Sanders, Scoville,  \&
  Soifer 1991;  Solomon et al.   1997), while luminous  transitions of
  heavy rotor molecules  such as CS J=3--2, and  HCN, HCO$^{+}$ J=1--0
  have demonstrated that, quite unlike typical quiescent spirals, most
  of the molecular gas in  this spectacular merger is very dense, with
  $\rm n\ga 10^5\,cm^{-3}$ (Solomon et al.  1990, 1992).  Such studies
  have recently culminated in the most extensive molecular line survey
  ever  conducted  for  such  objects  (Greve et  al.   2009),  making
  Arp\,220 a galaxy with the  best studied molecular gas reservoirs in
  the local Universe.  High resolution CO J=1--0, 2--1 interferometric
  imaging   reveals  two  compact   (2r$\la  $0.3$''$;   108\,pc)  gas
  concentrations  $\sim  $1$''$  (360\,pc)  apart (Scoville,  Yun,  \&
  Bryant 1997; Downes \& Solomon  1998; Sakamoto et al.  1999; Eckart,
  \& Downes 2001), whose intense starbursts dominate the IR luminosity
  of the  entire system.   Finally a major  advance was  recently made
  with  the  imaging of  the  CO  J=3--2  emission (and  the  adjacent
  continua at 860$\mu $m) of the warm and dense gas, that also implied
  substantial dust optical depths  even at submm wavelengths (Sakamoto
  et al. 2009).

We  report single  dish CO  J=6--5 observations  of Arp\,220  with the
James Clerk  Maxwell Telescope as  part of a  multi-J CO and  HCN line
survey  of LIRGs  ($\rm L_{IR}$$\ga  $ 10$^{11}$\,L$_{\odot}$)  in the
local Universe.  We demonstrate that an optically thick dust continuum
at  submm  wavelengths  submerges   this  CO  line  to  near-blackbody
emission, and  strongly affects the  emergent CO Spectral  Line Energy
Distribution (SLED)  and the C$^+$ fine-structure  line luminosity for
this  extreme  starburst.   Finally  we discuss  whether  similar  ISM
conditions are  present in dusty  starbursts in the  distant Universe,
lowering their  observed (high-J)/(low-J)  CO ratios well  below those
typical  of  star-forming  molecular  gas,  and  thus  limiting  their
diagnostic utility.

\section{Observations, line and continuum flux estimates}

We used the upgraded  dual-channel W/D (620--710\,GHz) receiver at the
James Clerk Maxwell  Telescope (JCMT)\footnote{The James Clerk Maxwell
Telescope is operated  by The Joint Astronomy Centre  on behalf of the
Science and  Technology Facilities Council of the  United Kingdom, the
Netherlands  Organisation for  Scientific Research,  and  the National
Research Council  of Canada.}  at  4092\,m altitude atop Mauna  Kea in
Hawaii, operating single sideband (SSB), to observe the CO J=6--5 line
($\rm \nu _{rest}$=691.473\,GHz) in  Arp\,220 on March 15th 2009 under
dry conditions  ($\rm \tau_{225\,GHz}$$\sim $0.04$-$0.06).   To ensure
the flattest  baselines possible,  but also measure  the corresponding
dust continuum at $\rm \lambda _{obs}$=442$\mu $m, we used the fastest
beam  switching mode  available (continuum  mode) with  a  beam switch
frequency of $\rm f_{bmsw}$=4\,Hz and  a throw of $30''$ (in azimuth).
The                                                               ACSIS
spectrometer\footnote{http://www.jach.hawaii.edu/JCMT/spectral\_line/Backends/acsis/acsis\-guide.html}
was used  at its widest  mode of 1.8\,GHz  ($\sim$780\,km\,s$^{-1}$ at
690\,GHz) and two separate tunings, yielding an effective bandwidth of
$\rm 3.235\,GHz  $ ($\sim $1400\,km\,s$^{-1}$). This  was necessary in
order   to  cover   the  widest   known  CO   line  in   local  ULIRGs
(FWZI$\sim$900--1000\,km\,s$^{-1}$).  The  typical system temperatures
were  $\rm  T_{sys}$$\sim $(2000$-$3000)\,K,  with  a  median of  $\rm
T_{sys}$$\sim $2500\,K  (including atmospheric absorption).   The beam
size  at 691\,GHz is  $\rm \Theta  _{HPBW}$=8$''$. Good  pointing with
such narrow beams is crucial and was checked every 45\,mins using both
absolute  W/D  and  differential   pointing  with  the  A\,3  receiver
(230\,GHz),  yielding $\rm  \sigma _x$$\sim  $$\sigma_y$$\sim $1.4$''$
(rms) (Figure  1).  The  aperture efficiency at  690\,GHz is  $\eta ^*
_a$=0.32,  estimated  from  the  Ruze  formula for  an  80\%  membrane
transmission, an rms dish surface  accuracy of $\sim $25$\mu $m, and a
$\rm \eta ^* _{a,0}$=0.68 (aperture efficiency for a perfect dish with
the  illumination  taper  applied  at  the JCMT),  and  verified  with
observations of Venus.  The  flux calibration uncertainty is estimated
with repeated  observations of compact spectral line  standards and is
$\sim 25\%$.

  Individual  spectra  were examined,  edited  for  bad channels,  and
co-added  (both W/D  channels) to  yield the  final spectrum  shown in
Figure 2 overlaid  with CO 3--2 (JCMT), and the  HCN J=1--0, CS J=3--2
lines obtained with the IRAM 30-m telescope (from Greve et al.  2009).
The  agreement between overall  FWZI's and  line centers  is excellent
although the  CO J=6--5 line  becomes weak towards the  velocity range
where  the  high density  tracer  CS  J=3--2  line ($\rm  n_{cr}$$\sim
$2$\times$10$^5$\,cm$^{-3}$) becomes  especially strong (see discusion
in section 3).  The integrated CO J=6--5 line flux is estimated from

\begin{equation}
\rm \rm S_{line}=\int _{\Delta V} S_{\nu } dV = \frac{8 k_B}{\eta ^{*}
_a  \pi  D^2} G(\sigma) \int   _{\Delta  V}  T^*  _A  dV=  \frac{\Gamma
(Jy/K)}{\eta ^{*} _a} G(\sigma) \int _{\Delta V} T^* _A dV,
\end{equation}

\noindent
where $\rm \Gamma _{JCMT} = 15.62(Jy/K)$, and

\begin{equation}
\rm G(\sigma)=1+8\,ln2\left(\frac{\sigma}{\Theta _{HPBW}}\right)^2,
\end{equation}

\noindent
with  $ \sigma  $$  \sim  $$\rm \sigma  _x$$\sim  $$\rm \sigma_y$  the
pointing residuals.   The aforementioned factor corrects  for the flux
bias resulting  when point-like sources are observed  with single dish
telescopes  with given  pointing rms  errors (Condon  2001).   This is
usually  insignificant  for   modern  mm/submm  telescopes  (e.g.   at
345\,GHz and the  same pointing rms error: G$\sim  $1.064), yet it can
become significant  at very high  frequencies, even for  the excellent
tracking  and pointing  of the  enclosed JCMT.   For  our observations
G=1.17, and we use it to scale our measured CO J=6--5 line and 442$\mu
$m    dust   continuum   fluxes    accordingly.    We    obtain   $\rm
S_{co}$(6$-$5)=(1170$\pm  $341)\,Jy\,km\,s$^{-1}$,  and  $\rm  S^{(c)}
_{442\,\mu m}$=(3.71$\pm  $0.96)\,Jy (using the line-free  part of the
spectrum and the  scaling factor in Equation 1),  or equivalently $\rm
S^{(c)}   _{434\,\mu   m}$=(3.85$\pm   $0.99)\,Jy  (scaled   by   $\rm
S_{\nu}$$\propto  $$\nu ^2$).   Regarding  the continuum  it is  worth
mentioning that  Arp\,220 is a  submm calibration source for  SCUBA at
the
JCMT\footnote{http://www.jach.hawaii.edu/JCMT/continuum/calibration/sens/potentialcalibrators.html},
where multiple images show a  compact source at 450$\mu $m ($\la 5''$)
with $\rm  S_{450\,\mu m}$=(2.77$\pm $0.06)\,Jy.  The  higher value of
$\rm  S_{450\,\mu m}$=(6.3$\pm  $0.8)\,Jy reported  by Dunne  \& Eales
(2001) is  thus  likely affected  by  calibration  error (Dunne  2009,
private  communication).   Finally  the  CO  J=6--5 flux  is  in  good
agreement with  that reported recently  by Matsushita et al.   2009 of
S$_{co}$(6--5)=(1250$\pm      $250)Jy\,km\,s$^{-1}$      using     the
SMA\footnote{A  fringe  pattern present  on  the  short baselines  has
likely corrupted the SMA channel map spectrum (Fig. 7 in Matsushita et
al.  2009),  and thus  the SMA spectrum  is not reproduced  here.  The
reported SMA line flux has  been corrected for that effect (Matsushita
2009,  private communication)}.   For  the purposes  of  this work  we
adopt: $\rm S_{co}$(6$-$5)=(1210$\pm $240)\,Jy\,km\,s$^{-1}$ (average:
JCMT  and  SMA),   $\rm  S^{(c)}  _{434\,\mu  m}$=(3.0$\pm  $0.06)\,Jy
(average: W/D, SMA, SCUBA calibrator database).

\section{The state of molecular gas and dust in Arp\,220}

A  large multi-J  CO, HCN,  HCO$^+$  and CS  line survey  by Greve  et
al. (2009)  makes Arp\,220 the  ULIRG with the  best-studied molecular
gas properties, and allows us to  place the measured CO J=6--5 line in
the best  possible perspective.  From  that study it  becomes apparent
that densities for the {\it bulk} of the molecular gas in Arp\,220 are
very high:  $\rm n(H_2)$$\ga  $10$^{5-6}$\,cm$^{-3}$, and thus  the CO
J=6--5  transition   is  expected   to  be  fully   thermalized  ($\rm
n_{crit}$$\sim  $ 5.8$\times $10$^4$\,cm$^{-3}$).   On the  other hand
global dust  continuum SEDs (e.g.   Lisenfeld, Isaak, \&  Hills 2000),
and  high resolution mm/submm  imaging (e.g.   Downes \&  Eckart 2007;
Sakamoto  et   al.   2008)  yield  typically   $\rm  T_{dust}$$\sim  $
(65--120)\,K.   For  the  high  gas  densities in  this  system:  $\rm
T_{kin}$$\sim  $$\rm T_{dust}$,  and the  CO line  emission  should be
peaking          at          $\rm         J_{max}$$\sim          $$\rm
\left[2k_BT_{kin}/E_{\circ}\right]^{1/2}$$\sim       $5--7       ($\rm
E_{\circ}/k_B$$\sim  $5.5\,K).  Clearly  the J=6--5  transition should
not only be  luminous but quite possibly the most  luminous CO line in
this  system,  with  typical  brightness temperature  ratios  of  $\rm
R_{65/J+1,J}$$\sim  $0.80--0.95  for  J+1=1--5,  even for  the  lowest
possible $\rm T_k$=65\,K and $\rm n(H_2)$$=$$\rm 10^5\,cm^{-3}$.

\subsection{High dust optical depths at submm wavelengths}

 Our   measured   CO   (6--5)/(1--0)  and   (6--5)/(3--2)   brightness
temperature ratios for  Arp\,220 are: $\rm R_{65/10}$=0.080$\pm $0.017
and  $\rm R_{65/32}$=0.082$\pm  $0.019  (CO J=1--0,  3--2 fluxes  from
Greve et  al 2009),  suggesting impossibly low  excitation conditions,
very much  incompatible with the  average properties of  its molecular
gas reservoir.   High dust optical  depths at short  submm wavelengths
provide the most  plausible explanation for the faint  CO J=6--5 line,
immersing  it   in  a  strong   nearly  blackbody  (and   thus  almost
featureless)  dust  continuum.   Such  large optical  depths  of  dust
continuum at  far-IR wavelengths were  first proposed for  Arp\,220 by
Lisenfeld et al.  (2000) where  $\rm \tau _{100}$$\sim $5--12 has been
deduced from  its far-IR/submm dust SED, with  Gonzalez-Alfonso et al.
(2004) arriving  at similar  conclusions using  a full  {\it  ISO} LWS
spectrum.  This has culminated with the recent work of Sakamoto et al.
2008 where large optical depths are reported even at submm wavelengths
($\rm  \tau_{860}$$\sim $1).   Their effect  on the  emergent  CO line
ratios  can be  easily shown  using a  simple model  of  an isothermal
mixture   of   molecular   gas    and   dust   in   LTE,   where   the
continuum-subtracted  source   line  brightness  temperature~is  (e.g.
Rohlfs \& Wilson 1996):

\begin{equation}
\rm \Delta T_{b,co}=e^{-\tau _d}\left[J(\nu_{co}, T)-J(\nu_{co}, T_{CMB})\right]
\left(1-e^{-\tau _{co}}\right),
\end{equation}

\noindent
where          $\rm          J(\nu,         T)$=$\rm          h\nu/k_B
 \left[exp(h\nu/k_BT)-1\right]^{-1}$,  and $\rm  \tau _{d}$  and $\tau
 _{co}$ are  the dust continuum and  CO line optical  depths.  Here is
 important  to  point out  that  the  equation  above assumes  a  form
 identical to  that for  line absorption by  a foreground  dust screen
 only  for  an  {\it  isothermal}  mixture of  line-emitting  gas  and
 continuum-emitting dust (i.e.   identical source functions).  In such
 a mixture  the high  dust optical depths  do not actually  reduce the
 emergent  line  strength  (as   they  would  for  a  dust  absorption
 ``screen'')  rather  than  make  both  continuum  and  spectral  line
 emission rise  up to a  common blackbody SED, with  Equation~3 simply
 expressing  the  diminishing  line-continuum  contrast.  For  a  dust
 emissivity  law $\rm  \tau _d(\nu)$$\propto  $$\nu  ^{\beta}$ ($\beta
 $=1-2), the brightness temperature ratio CO(6--5)/(J+1--J) transition
 would then be

\begin{equation}
\rm R^{(obs)} _{65/J+1,J}=
exp\left[-\tau_d(\nu _{J+1,J}) \left(\left(\frac{\nu _{65}}{\nu _{J+1,J}}\right)^{\beta}-1\right)\right]
\times R^{(int)} _{65/J+1,J},
\end{equation}

\noindent
where  $\rm R^{(int)} _{65/J+1,J}  $ is  the ``intrinsic''  line ratio
corresponding to $\rm \tau_d$=0.  For CO J=3--2 (also tracing the warm
star forming molecular  gas as the J=6--5 line)  and its corresponding
continuum emission Sakamoto et  al.  (2008) has deduced $\tau _{d}(\nu
_{32})$$\sim $1.   For $\rm n(H_2)$=$\rm 10^5\,cm^{-3}$  and a typical
$\rm  T_{kin}$=(65--90)\,K  our  Large  Velocity Gradient  (LVG)  code
yields  $\rm R^{(int)}  _{65/32}$$\sim  $0.85--0.90.  Inserting  these
values   into   Equation~4   yields  $\rm   R^{(obs)}   _{65/32}$$\sim
$0.042--0.32 (for $\beta $=1-2), comfortably encompassing the observed
value, and demonstrating how substantial submm dust optical depths can
suppress (high-J)/(low-J)  CO line ratios.  The  larger suppression of
CO J=6--5 in  the eastern nucleus of Arp\,220,  contributing mostly to
the ``horn'' at $\sim $5600\,km\,s$^{-1}$  of the CO line profiles, is
then  a  likely   result  of  the  larger  gas   densities  (and  thus
correspondingly  larger   dust  optical  depths).    This  is  clearly
indicated by  the rising  CS J=3--2 line  emission (Figure 2)  and the
larger   HCN(4--3)/CO(3--2)  ratio  (Figure   3)  towards   $\rm  \sim
$(5500--5600)\,$\rm km\,s^{-1}$ than towards $\sim $(5300--5400)\,$\rm
km\,s^{-1}$, the latter range associated with the western nucleus.

Thus the CO  J=6--5 line (and partially even J=3--2)  is affected by a
dust continuum  rising to an  almost black body  curve from IR  out to
short submm  wavelengths.  This is  further indicated by  the observed
ratio $\rm  S_{434\mu m}/S_{860\mu  m}$$\sim $5.45$\pm $0.82  (for the
two nuclei: $\rm S_{860\mu m}$=(0.55$\pm $0.082)\,Jy, from Sakamoto et
al.    2008),  which   is   close   to  the   black   body  value   of
$(860/434)^2$=3.93. Indeed from

\begin{equation}
\rm \frac{S_{434\mu m}}{S_{860\mu m}}=\left(\frac{\nu _{434}}{\nu _{860}}\right)^2
\frac{1-exp\left[-\left(\frac{\nu _{434}}{\nu _{860}}\right)^{\beta}\,\tau_{860}\right]}
{1-exp\left(-\tau _{860}\right)},
\end{equation}

\noindent
and $\rm  S_{434\mu m}/S_{860\mu  m}$=5.45 we find  $\tau _{860}$$\sim
$1.25 (for $\beta $=2), while for $\rm S_{434\mu m}/S_{860\mu m}$=6.27
($+\sigma$ of  the measured value)  the minimum optical depth  is $\rm
\tau _{860}\sim $0.95.  Dust optical  depths large enough to bring all
this about are certainly present  in Arp\,220 (Sakamoto et al.  2008),
while  the globally  supressed J=6--5  line further  demonstrates that
much of the large molecular gas  mass in this system is ``cloaked'' by
these  high dust  optical depths  rather than  only  small sub-regions
around e.g.  an  AGN (Downes \& Eckart 2007).  If  this is typical for
ULIRGs,  it can  {\it modify  the emergent  CO SLEDs,  and  affect the
diagnostic  value of  molecular  lines at  high  frequencies for  such
extreme systems,} a possibility further discussed in Section 4.

Finally  we  note that  the  average state  of  the  molecular gas  in
Arp\,220, as indicated by the HCN, CS, and HCO$^+$ lines, is so highly
excited that  even CO J=6--5 line  fluxes up to $\sim  $3 times higher
would leave $\rm  R_{65/32}$ {\it well below what  is expected of such
gas and is typically observed in starbursts} (see Figure 5 in section
4.2), still suggesting substantial  dust optical depths at short submm
wavelengths surpressing high-J CO line emission.

\subsection{Arp\,220: the C$^+$ line luminosity deficit resolved}

For a $\rm \tau _{850}\ga 1$ ``cloaking'' the bulk of the gas and dust
in Arp\,220, and conservatively  assuming $\beta $=1, yields $\rm \tau
_{158}\ga 5.4$ at the emission  wavelength of the C$^+$ fine structure
line.  This  is more  than enough to  almost completely  suppress this
strong ISM cooling line into  a featureless black body dust continuum,
favoring large dust optical depths as the cause for its weakness among
the other  explanations proposed (e.g. Malhotra et  al. 1997).  Indeed
the faint CO J=6--5 line,  besides corroborating the high dust optical
depths at IR/submm  wavelengths in Arp\,220, also makes  them the most
likely  cause  of its  C$^+$  line  luminosity  ``deficit'' (found  in
ULIRGs, with Arp\,220 having the  largest; Luhman et al.  1998, 2003).
It does  so by  negating a prominent  alternative explanation  for the
C$^+$  line surpression,  namely  very dense  PDRs (where  precipitous
C$^+$ recombination would remove it  from the ISM). This is because CO
J=6--5 is a different spectral line (governed by different physics and
chemistry than C$^+$)  and one that {\it ought} to  be luminous if that
alternative  explanation held.   Indeed,  for dense  PDRs immersed  in
strong far-UV radiation fields high-J CO lines are expected to be very
luminous  and  an  alternative  cooling  ``channel'' to  that  of  the
suppressed C$^+$ line, balancing  the tremendeous ISM heating expected
in ULIRGs (Papadopoulos, Isaak, \& van der Werf~2007). Finally it must
be noted that  if high dust optical depths  at short submm wavelengths
are responsible for surpressing both the high-J CO and the C$^+$ lines
in ULIRGs, it follows  that the starburst systems with dust-surpressed
(high-J)/(low-J)  CO  line  ratios  will  also  be  those  with  small
L(C$^+$)/L$\rm _{IR}$.

\section{High dust optical depths in ULIRGs: from the IR to the submm}

The  discovery of  dust continuum  optical depths  that in  ULIRGs can
remain  substantial  even  out  to  short  submm  wavelengths  is  the
culmination of a series  of studies successively ``pushing'' the $\tau
_{\lambda }\ga  1$ limit longwards in wavelength  for these remarkable
systems  (Condon  et al  1991;  Solomon  et  al.  1997;  Lisenfeld  et
al. 2000;  Sakamoto et al.  2008).   Condon et al.  were  the first to
point out that high dust extinction at far-IR wavelengths is needed to
explain radio  continuum sizes of  ULIRGs that are smaller  than their
minimum IR  black-body emission sizes.  They  argued that re-radiation
of the  IR light from a  compact starburst by  dust ``layers'' further
out that remain optically thick at far-IR wavelengths can yield larger
effective  IR-continuum source  sizes  than the  true starburst  sizes
(revealed   by   radio  continuum).    For   the   compact  gas   disk
configurations in  ULIRGs (expected  in mergers and  their dissipative
gas  motions)  such  large  dust  optical  depths  can  make  both  IR
``colors''   and  ISM   line   ratios  at   short  submm   wavelengths
viewing-angle   dependent   and    thus   much   {\it   reduce   their
AGN-vs-starburst diagnostic  power.}  This is  indeed the case  for IR
colors where ``IR-cool'' ULIRGs may  not be lacking an AGN but instead
host AGN that are heavily  obscured along the particular line of sight
observed  (Condon et  al.  1991),  while in  nearly  face-on ULIRG/QSO
systems (e.g.   Mrk\,231) such AGN  remain visible with their  warm IR
colors discernible through smaller  dust optical depths.  It now seems
that molecular line diagnostics  at short submm wavelengths may suffer
from similar effects.

\subsection{The diagnostic power of far-IR/submm ISM lines revisited}

A  general  effect  of  high   dust  optical  depths  at  short  submm
wavelengths on  the CO SLEDs would  be to make  them appear ``cooler''
than  those typical  for  star-forming gas.   Thus dust-suppressed  CO
(high-J)/(low-J) ratios can  be hard to discern from  those typical of
genuinely low-excitation~gas.  High dust optical depths at short submm
wavelengths in  ULIRGs are particularly troubling for  ISM lines whose
physics  makes  them  diagnostic   of  deeply  ``buried''  AGN  versus
starbursts as  the energy source of their  prodigious IR luminosities.
This is especially true for line frequencies of $\rm \nu \ga$~690\,GHz
whose  ratios, when  unaffected  by dust  absorption,  offer the  best
discrimination between the various  ISM excitation mechanisms, such as
X-rays from a deeply buried  AGN (Meijerink \& Spaans 2005), or far-UV
photons from their starbursts (Meijerink, Spaans, \& Israel 2006).  In
such cases  ratios of lines  adjacent in frequency  (e.g.  HCO$^+$/HCN
for the same  rotational level) will have to  be used as AGN/starburst
discriminators.   Finally, high  dust  optical depths  at short  submm
wavelengths can  affect molecular line observations of  ULIRGs at high
frequencies with  the upcoming {\it Herschel  Space Observatory.}  For
Arp\,220 in  particular faint and apparently  sub-thermally excited CO
lines are to be expected at  $\nu $$\ga $690\,GHz (Figure 4), and this
will  also affect  high-J  transitions of  any  molecules tracing  the
star-forming gas phase (e.g. HCN).

\subsection{Evidence for high dust optical depths at  submm wavelengths in distant starbursts}

Low  (high-J)/(low-J) CO  line ratios  quite atypical  of star-forming
molecular gas have been measured in some of the most IR-luminous ($\rm
L_{IR}$$\sim   $$10^{13}$\,L$_{\odot}$)  starbursts  in   the  distant
Universe   (Tacconi   et  al.    2006),   a   Ly-break  galaxy   whose
CO(7--6)/(3--2) brightness temperature  ratio is $\rm R_{76/32}$=0.030
(Baker et al.  2004), and in a submm-bright ULIRG at z=1.44 where $\rm
R_{54/21}$$\sim  $0.16  (Papadopoulos \&  Ivison  2002).  However  the
wealth of  molecular line  data for Arp\,220  that allowed its  low CO
(6--5)/(3--2) ratio to  be attributed to large dust  optical depths at
short  submm  wavelengths  is  not  available yet  for  high  redshift
systems.  Thus, as discussed in the previous section, ``cool'' CO line
ratios can be due to either low-excitation gas or to high dust optical
depths.

    Dominant amounts of low-excitation  and SF-idle gas extending well
 beyond star-forming galactic nuclei are indeed well-known features of
 star-forming  spirals in the  local Universe  and even  of archetypal
 starbursts such as M\,82  (Weiss, Walter, \& Scoville~2004).  However
 recent high resolution CO  imaging of submm-bright galaxies (SMGs) at
 high  redshifts finds  no evidence  for such  low-excitation extended
 molecular gas reservoirs  (Tacconi et al.  2006; Iono  et al.  2009),
 yet it  also measures CO (6--5)/(3--2), (7--6)/(3--2)  line ratios of
 $\la 0.35$ in many such systems.   In one extreme such case, the most
 compact  SMG (SMM\,J044307+0210 at  z=2.5) has  the lowest  such line
 ratio  of   $\rm  R_{76/32}$$\sim  $0.13  (Tacconi   et  al.   2006).
 Indicatively, for  nearby star-forming  galactic nuclei such  CO line
 ratios are  $\sim $0.8--0.9 (Mao et  al.  2000; Bayet  et al.  2006),
 and values  of $\la $0.30 are typical  of low-excitation SF-quiescent
 gas. In Figure 5 we  show the $\rm R_{65/32}$ ratios for star-forming
 and  quiescent environments in  the local  Universe along  with those
 available  for  systems at  high  redshifts.   From  this it  becomes
 obvious that:  a) nearby  starburst environments behave  as expected,
 tracing  the  entire range  of  $\rm  R_{65/32}$  values typical  for
 star-forming molecular gas, and b) several high luminosity SMGs ($\rm
 L_{IR}$$>$$10^{12}$\,L$_{\odot}$)     fall     well    below     this
 range,\footnote{If  a  ``neighboring''  CO  line  ratio  (e.g.   $\rm
 R_{76/32}$)  rather than $\rm  R_{65/32}$ is  avalaible for  a high-z
 system, we obtain  the average value and range  of $\rm R_{65/32}$ by
 using the  available one to  find compatible LVG solutions  over $\rm
 T_k$=(30--110)\,K,                    $\rm                    \langle
 n(H_2)\rangle$=($10^2$--$10^7$)\,cm$^{-3}$,    assuming   also   $\rm
 K_{vir}$$\sim  $1.}  and  well into  that typical  for low-excitation
 SF-idle environments.  In some  SMGs $\rm R_{65/32}$ falls even below
 the  minimum value of  $\sim $0.3,  expected for  the dense  and cold
 starless cores with minimum temperatures of $\rm T_{kin}$$\sim $10\,K
 (set  by  Galactic  cosmic  rays),  and  densities  of  n(H$_2$)$\sim
 $10$^5$\,cm$^{-3}$.  Indicatively $\rm  R_{65/32}$ for Arp\,220 would
 remain below that value even for  a CO J=6--5 line flux that is $\sim
 $3 times higher than the one observed.

 In order to  better quantify this we conduct  Large Velocity Gradient
 (LVG) radiative transfer modeling of  the CO line ratios reported for
 SMGs by Tacconi et al.   (2008).  For $\rm R_{65/32}\la 0.3$ and $\rm
 R_{76/32}\la    0.15$   we    find   no    solutions    within   $\rm
 T_{kin}$$\sim$(40--90)\,K, $\rm  n(H_2)$ $\ga $$10^4$\,cm$^{-3}$ {\it
 and}  $\rm   K_{vir}$$\sim  $1  (self-gravitating   gas  phase),  the
 parameter  space  typical for  star-forming  gas.  The K$_{\rm  vir}$
 parameter is

\begin{equation}
\rm K_{vir}=\frac{\left(dV/dr\right)_{LVG}}{\left(dV/dr\right)_{virial}}\sim 
1.54\frac{[CO/H_2]}{\sqrt{\alpha}\Lambda _{co}}\left(\frac{n(H_2)}{10^3\,
 cm^{-3}}\right)^{-1/2},
\end{equation}

\noindent
($\alpha $$\sim  $1--2.5, depending on the cloud  density profile) and
quantifies the kinematics of the  gas responsible for the CO emission,
with $\rm K_{vir}$$\sim $1  (within factors of $\sim $2--3) indicating
a virialized gas  phase and $\rm K_{vir}$$\gg $1  an unbound one (e.g.
Greve et  al.  2009).   Only for $\rm  R_{76/32}$$\sim $0.3  are there
solutions      $\rm      T_{kin}$=[40\,K,      (65--80)\,K],      $\rm
n(H_2)$=$10^4$\,cm$^{-3}$, i.e.  typical  for star-forming gas, except
that $\rm K_{vir}$=[4, 11] indicates  unbound gas motions.  The low CO
(high-J)/(low-J) line ratios found in some distant starbursts are thus
{\it  not compatible  with the  typical conditions  of warm  and dense
star-forming gas.}  Hence, unless one is willing to postulate dominant
amounts of  cold and/or  diffuse gas in  some of the  most spectacular
starbursts  in the Universe,  high dust  optical depths  at rest-frame
submm  wavelengths  are  the   only  other  way  of  surpressing  such
(high-J)/(low-J)  CO line  ratios.  Finally,  it is  unlikely Arp\,220
will be the only ULIRG, near  or far, where this happens making its CO
SLED appear ``cool'' and atypical of its vigorously star-forming dense
gas.

In  summary high  dust optical  depths at  short submm  wavelengths in
extreme starbursts could then seriously impact

\begin{itemize}

\item  the detectability  of  CO J+1$\rightarrow$J,  J+1$>$6 lines  of
       starbursts in  the local or distant Universe  with Herschel and
       ALMA respectively

\item  the  diagnostic power  of  {\it  high-frequency} molecular  and
      atomic line  ratios, (e.g.  deduced  densities and temperatures,
      AGN-induced XDRs versus starburst-related PDRs)

\item the  thermal balance in the  molecular ISM of  ULIRGs (with $\rm
       C^+$  and high-J  CO  line cooling  diminished, dust  continuum
       maybe the dominant cooling ``channel'')

\end{itemize}

\noindent
We  will further  explore these  issues in  a future  paper  that will
 include the  entire suite of CO  lines observed for the  LIRGs in our
 sample  and  their  corresponding molecular SLEDs  for  heavily  dust
 obscured environments.

\section{Conclusions}

We report measurements  of the high-excitation CO J=6--5  line and its
adjacent  dust  continuum at  442$\mu  $m  of  the prototypical  Ultra
Luminous  Infrared  Galaxy  Arp\,220  with  the  James  Clerk  Maxwell
Telescope in Hawaii. Our findings can be summarised as~follows:

\begin{enumerate}

\item  The  CO J=6--5  line  is  very  faint, with  CO  (6--5)/(1--0),
      (6--5)/(3--2)    brightness   temperature    ratios    of   $\rm
      R_{65/10}$$\sim  $$\rm R_{65/32}$$\sim  $0.08,  much lower  than
      those expected  from the  warm and dense  star-forming molecular
      gas that is present in this extreme starburst.

\item High dust optical depths  with $\rm \tau (\nu \ga 350\,GHz)$$\ga
      $1 are responsible for the  faint CO J=6--5 line in Arp\,220 and
      will strongly  modify its emergent  CO SLED. We  anticipate that
      for  this ULIRG  the now  spaceborne Herschel  Space Observatory
      will  observe  faint  and  apparently sub-thermally  excited  CO
      lines beyond $\sim $690\,GHz.

\item The C$^+$ and the CO J=6--5 line luminosity ``deficits'' in this
      system  have  the same  cause:  high  far-IR/submm dust  optical
      depths.

\item The  low CO (high-J)/(low-J)  line ratios found also  in several
      starbursts  at high  redshifts could  also be  due to  high dust
      optical depths at short sub-mm wavelengths, though in individual
      objects this can be hard to distinguish from large reservoirs of
      genuinely low-excitation  molecular gas.

\item High dust optical  depths at far-IR/submm wavelengths can affect
       the  diagnostic power  of  molecular and  atomic  lines in  the
       spectral regime where it is expected to be the greatest, making
       their  intensity ratios depend  on differential  absorption and
       viewing angle rather than underlying gas excitation conditions.

\end{enumerate}

\acknowledgments We  would like  to thank Telescope  System Specialist
Jim Hoge  as well  as the  entire supporting team  of the  James Clerk
Maxwell Telescope for conducting a superb set of measurements on March
15th  2009 that made  this work  possible. Moreover  we would  like to
thank Dr Satoki Matsushita for informing us about some issues with the
SMA  data,  and  Dr  Loretta  Dunne for  discussions  regarding  SCUBA
calibration  issues.  PPP  would  like  to thank  Dr  Xilouris of  the
National Observatory of Athens for helpful discussions, and assistance
with  gathering/plotting literature  data.  Finally  we would  like to
thank the anonymous referee for his/hers pointed and clear suggestions
that led to critical improvements of the original manuscript.
\vspace*{2.0cm}

{\it   NOTE-ADDED-IN-PROOF:}  Recent   results  from   Herschel  Space
Observatory  suggest  a CO  J=6--5  line-integrated  flux 2-2.5  times
higher than that measured by the JCMT and the SMA. Pending solution of
various calibration issues with the Herschel FTS instrument, we note
that such values still leave our main conclusions intact (section 3.1)

{}

\clearpage

\begin{figure}
\epsscale{1.1}
\plotone{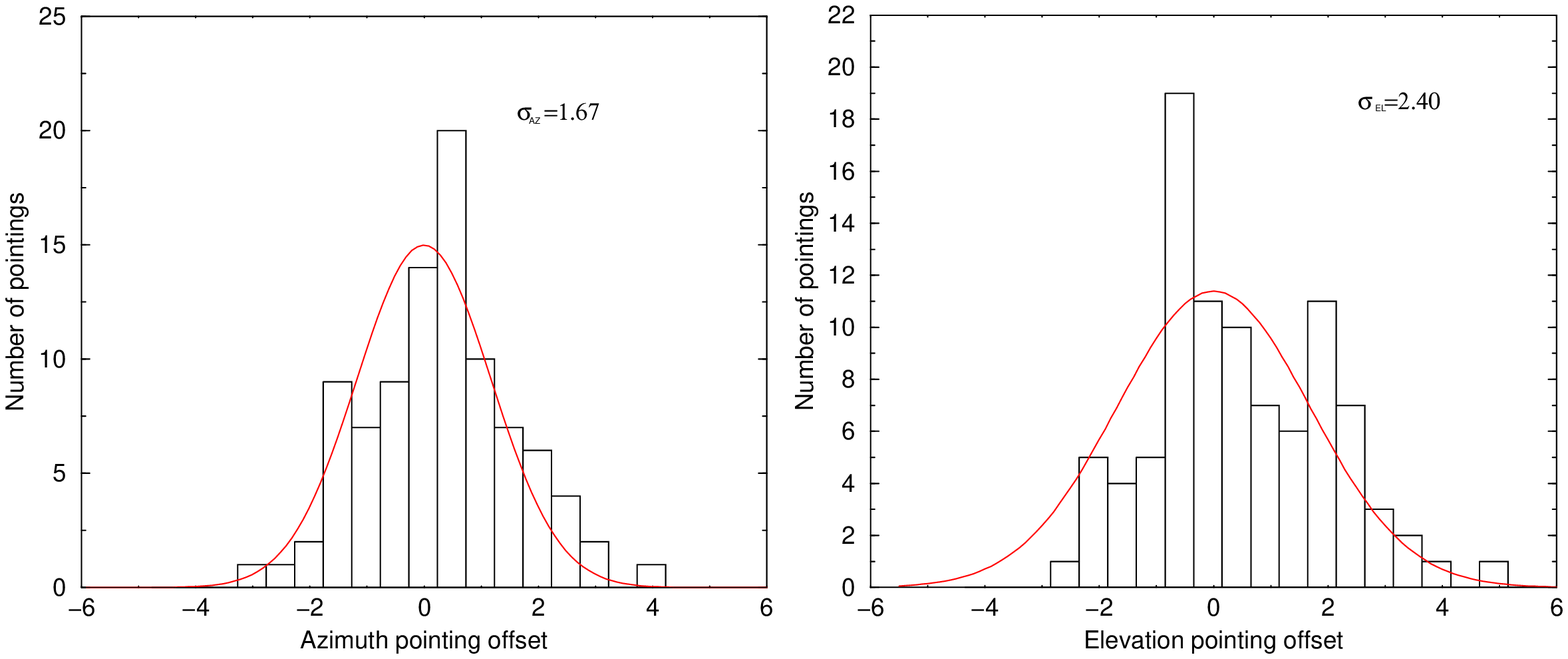}
\caption{The distributions of pointing offsets obtained during all W/D
observing periods.   Values with  $\rm |\sigma|$$\la $2$''$  are those
that  apply to  source observations,  with any  larger  ones typically
obtained  after  large  changes  in  azimuth and  elevation  when  the
telescope changed  sky sectors.  The pointing  offsets during Arp\,220
observations  are:   $\rm  \sigma  _{az}$$\sim   $$\sigma  _{el}$$\sim
$1.4$''$,  and  $\rm  \sigma   _r$=$\rm  (\sigma  ^2  _{az}+\sigma  ^2
_{el})^{1/2}$$ \sim $2$''$.}
\end{figure}

\clearpage

\newpage

\begin{figure}
\epsscale{1.2}
\plottwo{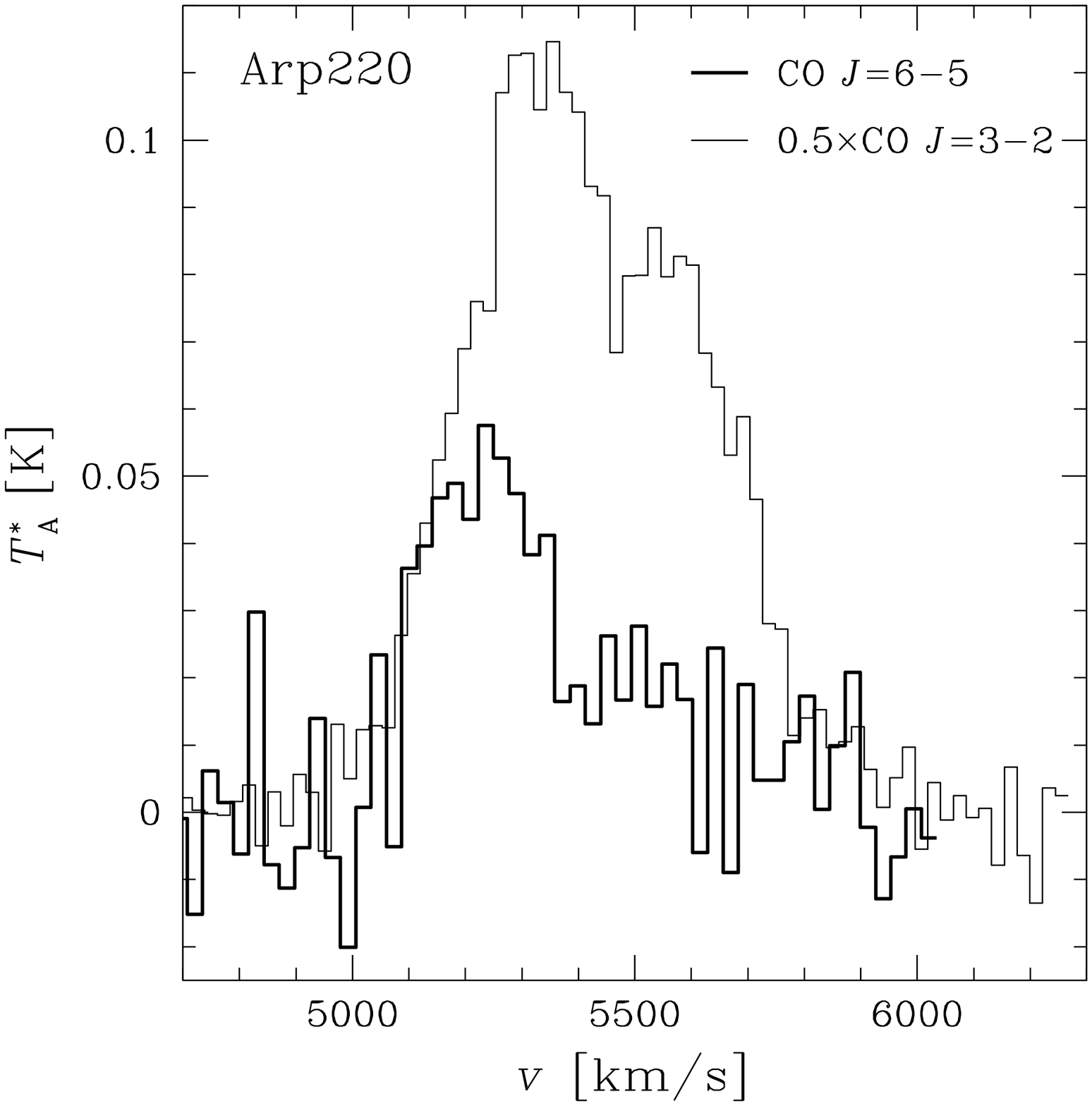}{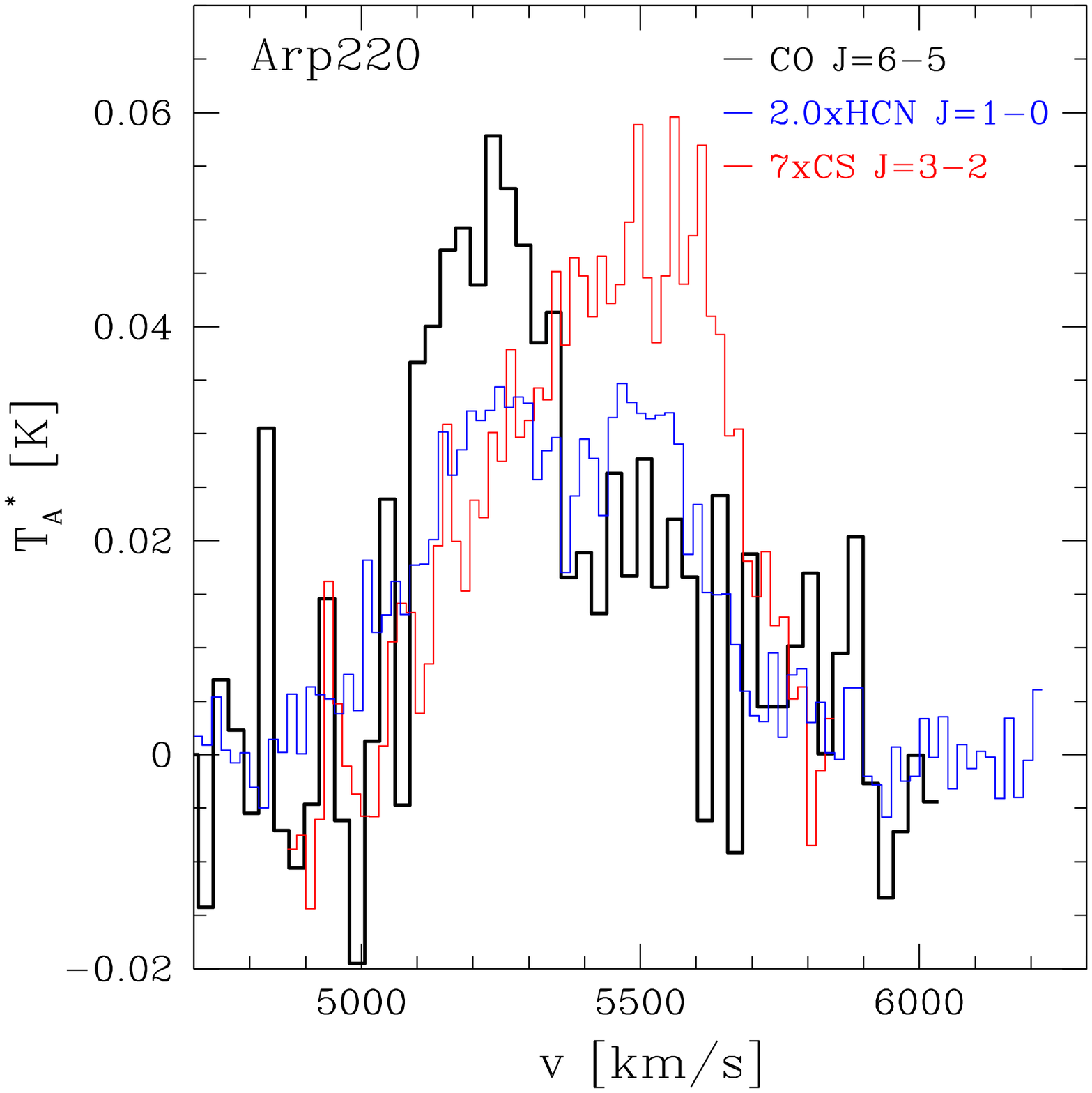}
\caption{Arp\,220: $\rm \alpha $=$\rm 15  ^h\ 34 ^{m}\ 57.24 ^s$, $\rm
\delta $=$+23 ^{\circ}\ 30 ^{'}\  11.2 ^{''}$ (J2000).  Top: CO J=6--5
(thick line), overlaid  to CO J=3--2 (thin line),  with resolutions of
$\rm  \Delta  V _{ch}$=27\,km\,s$^{-1}$  (J=6--5)  and  $\rm \Delta  V
_{ch}$=23\,km\,s$^{-1}$      (J=3--2),       and      centered      at
cz=5450\,km\,s$^{-1}$  (LSR).   The   thermal  rms  error  across  the
line-free part of the spectrum is $\rm \delta T^* _A$$\sim $12\,mK for
both lines. Bottom:  the CO J=6--5 line overlaid  with HCN J=1--0, and
CS J=3--2 lines,  obtained with the IRAM 30-m  telescope (adapted from
Greve et al. 2009).}

\end{figure}

\newpage

\begin{figure}
\epsscale{0.8}
\plotone{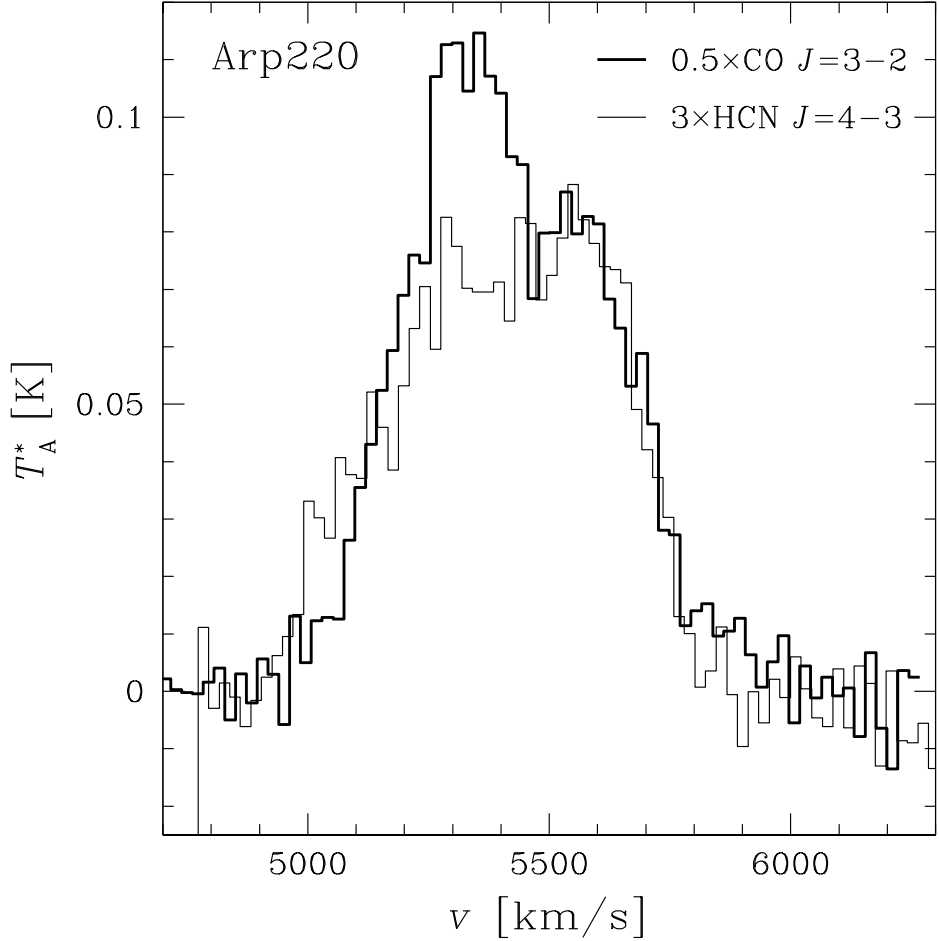}
\caption{The HCN J=4--3 (thin line) overlaid to CO J=3--2 (thick line)
  spectrum,     with     resolutions      of     $\rm     \Delta     V
  _{ch}$=22--23\,km\,s$^{-1}$ (HCN J=4--3, CO  J=3--2).  The CO and HCN
  emission of  the denser eastern  nucleus corresponds to  the feature
  centered  at $\sim $5500--5600\,km\,s$^{-1}$  where the  HCN/CO line
  ratio is the  largest (Greve et al.  2009).   The thermal rms errors
  across  the line-free  part  of  the spectra  are:  $\rm \delta  T^*
  _A$$\sim $2\,mK  (HCN J=4--3) and  $\rm \delta T^*  _A$$\sim $12\,mK
  (CO 3--2).}
\end{figure}

\newpage

\begin{figure}
\centering
\epsscale{1.0}
\plotone{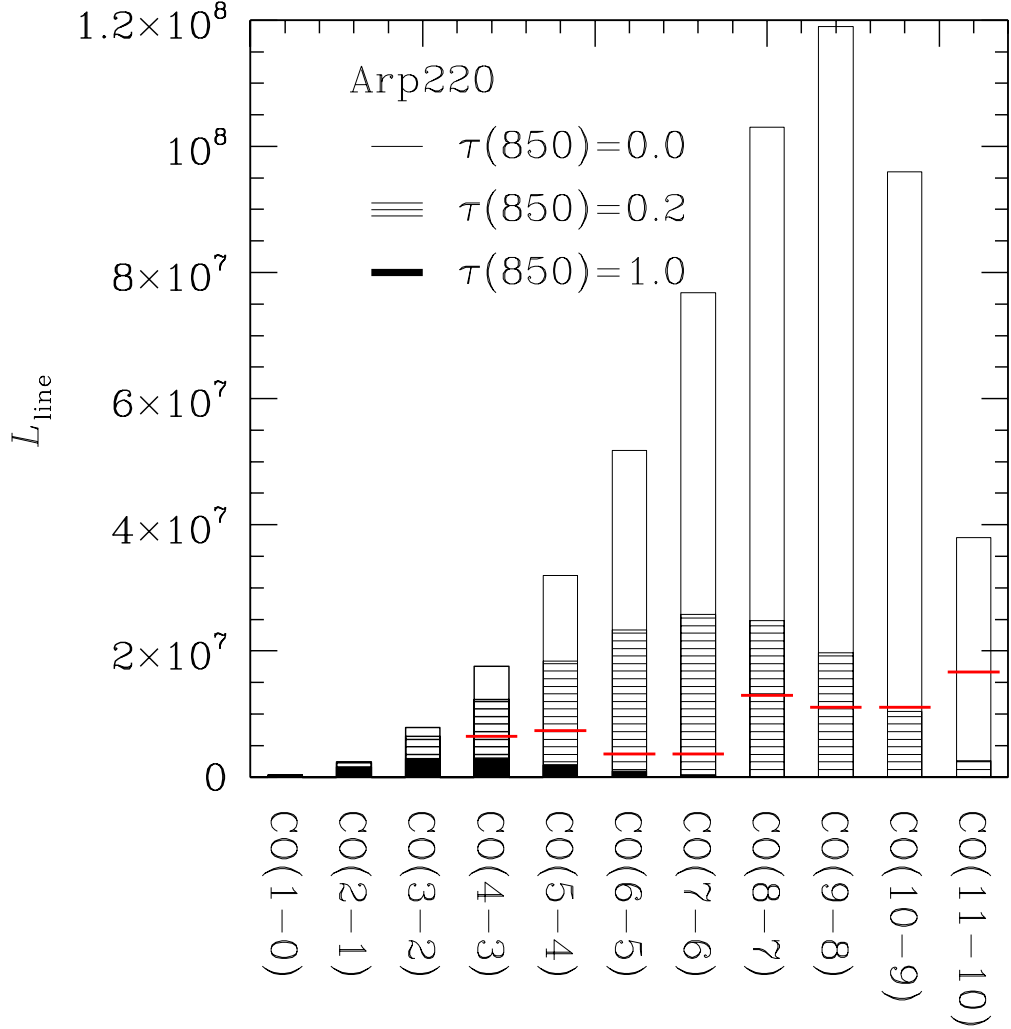}
\caption{The CO  SLED of Arp220 and  the effects of  dust (see Section
  4.1), where $L_{line}$ is given in solar luminosities.  The red bars
  denote  the  5sigma-1hr  detection   limits  for  the  SPIRE-FTS  on
  Herschel,   evaluated  using   the   observatory  integration   time
  calculator and assuming that all the line flux is contained in 1 (CO
  J=4--3 up to  J=7--6), 2 (CO J=8--7 up to  J=10--9) and 3 resolution
  elements (CO  J=11--10).  It  is clear that  CO lines  above J=6--5,
  7--6 can  be faint and  appear as sub-thermally excited,  while very
  high-J transitions such as J=10--9, 11--10 will be undetectable.}

\end{figure}

\newpage

\begin{figure}
\epsscale{1.0}
\plotone{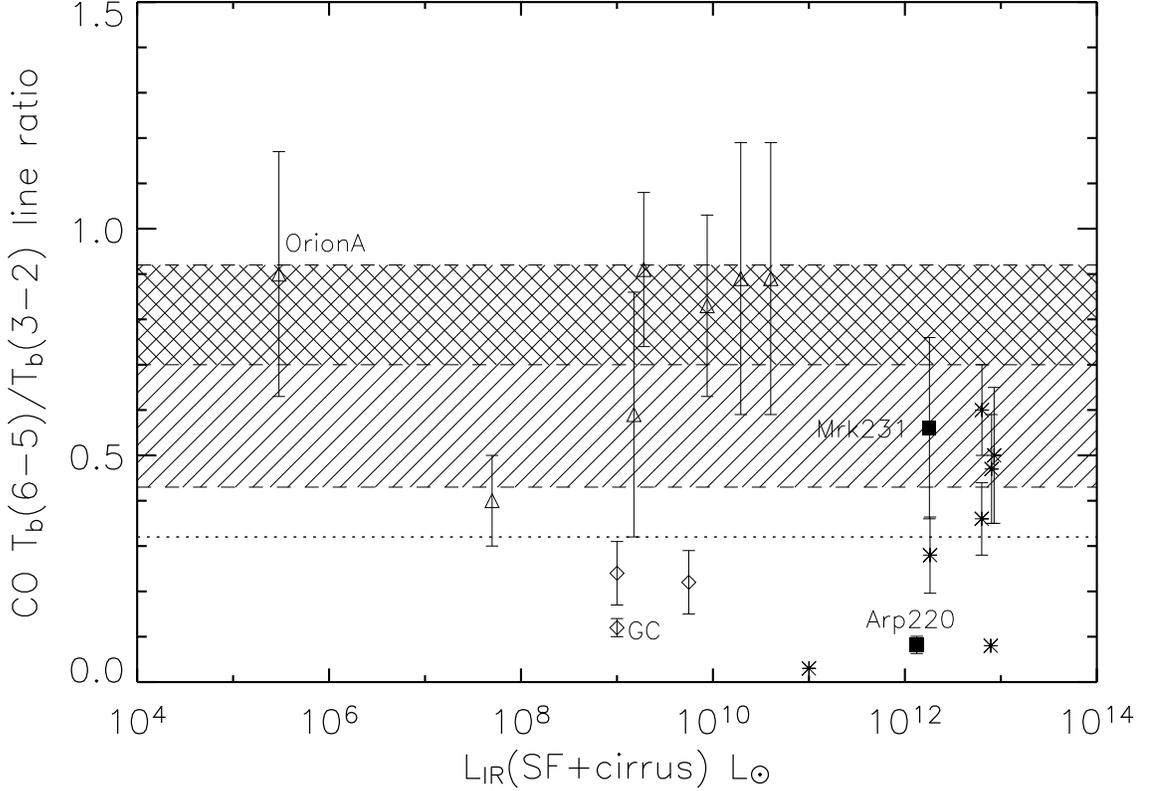}
\caption{The  CO (6--5)/(3--2)  ratio versus  L$_{\rm IR}$(40--400$\mu
 $m) for:  a) quiescent systems (diamonds),  nearby starburst galaxies
 or  starburst nuclei (triangles),  and c)  SMGs (stars).   The ULIRGs
 Arp\,220,  Mrk\,231, the  star-forming  region in  Orion\,A, and  the
 SF-quiescent Galactic  Center are also shown.  The  shaded areas mark
 the  $\rm R_{65/32}$  range  for typical  star-forming  gas in  LIRGs
 (hatched),  and  the  extreme  star-forming  environments  of  ULIRGs
 (cross-hatched).  The  first was estimated from LVG  models for: $\rm
 T_{kin}$=(30--100)\,K,  $\rm  n(H_2)$=($10^4$--$10^6$)\,cm$^{-3}$ and
 $\rm  K_{vir}$=1  (fully   encompassing  the  typical  conditions  of
 star-forming gas),  while the second narrower range  is obtained from
 LVG solutions constrained  by the HCN(3--2)/(1--0) average brightness
 temperature  ratio  of  $\langle \rm  R_{32/10}(HCN)\rangle$=0.55  of
 nearby ULIRGs (from  Krips et al.  2008; Gracia-Carpio  et a.  2008).
 The dotted line  marks the lowest possible $\rm  R_{65/32}$ for dense
 and   cold  GMC  cores   ($\rm  n(H_2)$$>   $10$^4$\,cm$^{-3}$,  $\rm
 T_{kin}$=10\,K) where both transitions would be fully thermalized and
 optically thick,  and which  is also the  limit below which  only gas
 with $\rm n(H_2)$$< $$10^4$\,cm$^{-3}$  can be found.  {\it CO data:}
 Galactic Center: Fixsen et al.   1999; Orion\,A: Marrone et al. 2004;
 Mrk\,231:  Papadopoulos  et al.   2007;  Arp\,220:  this work;  SMGs:
 Tacconi et al.  2006, Papadopoulos \& Ivison 2002; nearby starbursts:
 Wild et al.  1992, Mao et  al.  2000, Bradford et al.  2003, Bayet et
 al.~2006.}
\end{figure}

\end{document}